\documentclass[lettersize,journal,12p]{IEEEtran}
\usepackage{amsmath,amsfonts}
\usepackage{array}
\usepackage[caption=false,font=normalsize,labelfont=sf,textfont=sf]{subfig}
\usepackage{textcomp}
\usepackage{stfloats}
\usepackage{url}
\usepackage{graphicx}
\usepackage{cite}
\usepackage{hyperref}

\usepackage{pgfplots}
\usepackage{tikz}
\usetikzlibrary{calc}
\makeatletter
\newcommand{\gettikzxy}[3]{%
  \tikz@scan@one@point\pgfutil@firstofone#1\relax
  \edef#2{\the\pgf@x}%
  \edef#3{\the\pgf@y}%
}
\usetikzlibrary{spy,backgrounds}

\usepackage{comment}
 
\usepackage{acro}

\usepackage{xcolor}

\DeclareAcronym{aoa}{
short=AOA,
long= angle-of-arrival,
}
\DeclareAcronym{aod}{
short=AOD,
long= angle-of-departure,
}
\DeclareAcronym{ap}{
short=AP,
long= access point,
}
\DeclareAcronym{bs}{
short=BS,
long= base station,
}
\DeclareAcronym{d2d}{
short=D2D,
long= device-to-device,
}
\DeclareAcronym{hwi}{
short=HWI,
long= hardware impairment,
}
\DeclareAcronym{kpi}{
short=KPI,
long= {key performance indicator},
}
\DeclareAcronym{islac}{
short=ISLAC,
long= {integrated sensing, localization, and communication},
}
\DeclareAcronym{los}{
short=LOS,
long= line-of-sight,
}
\DeclareAcronym{las}{
short=L\&S,
long= localization and sensing,
}
\DeclareAcronym{lmf}{
short=LMF,
long= location management function,
}
\DeclareAcronym{amf}{
short=AMF,
long= access and mobility management function,
}
\DeclareAcronym{nb}{
short=NB,
long= narrowband,
}
\DeclareAcronym{peb}{
short=PEB,
long= position error bound,
}
\DeclareAcronym{ris}{
short=RIS,
long= reconfigurable intelligent surface,
}
\DeclareAcronym{rsu}{
short=RSU,
long= roadside unit,
}
\DeclareAcronym{siso}{
short=SISO,
long= single-input-single-output,
}
\DeclareAcronym{slam}{
short=SLAM,
long= simultaneous localization and mapping,
}
\DeclareAcronym{srs}{
short=SRS,
long= sounding reference signal,
}
\DeclareAcronym{trp}{
short=TRP,
long= transmission and reception point,
}
\DeclareAcronym{tx}{
short=TX,
long= transmitter,
}
\DeclareAcronym{rx}{
short=RX,
long= receiver,
}
\DeclareAcronym{risc}{
short=RISC,
long= RIS controller,
}
\DeclareAcronym{riso}{
short=RISO,
long= RIS orchestrator,
}
\DeclareAcronym{ue}{
short=UE,
long= user equipment,
}
\DeclareAcronym{v2x}{
short=V2X,
long= vehicle-to-everything,
}
\DeclareAcronym{wb}{
short=WB,
long= wideband,
}
\DeclareAcronym{xr}{
short=XR,
long= extended reality,
}
\DeclareAcronym{tdoa}{
short=TDOA,
long= time-difference-of-arrival,
}

\begin{document}
\bstctlcite{IEEEexample:BSTcontrol}

\title{RISs and Sidelink Communications in Smart Cities: The Key to Seamless Localization and Sensing}





\author{
Hui~Chen,~\IEEEmembership{Member,~IEEE},
Hyowon~Kim,~\IEEEmembership{Member,~IEEE},
Mustafa~Ammous,~\IEEEmembership{Student~Member,~IEEE}, 
Gonzalo~Seco-Granados,~\IEEEmembership{Senior~Member,~IEEE},
George~C.~Alexandropoulos,~\IEEEmembership{Senior~Member,~IEEE},
Shahrokh~Valaee,~\IEEEmembership{Fellow,~IEEE}, 
and~Henk~Wymeersch,~\IEEEmembership{Senior~Member,~IEEE}
}

\maketitle

\begin{abstract}
A smart city involves, among other elements, intelligent transportation, crowd monitoring, and digital twins, each of which requires information exchange via wireless communication links and localization of connected devices and passive objects (including people). Although localization and sensing (L\&S) are envisioned as core functions of future communication systems, they have inherently different demands in terms of infrastructure compared to communications.
Wireless communications generally requires a connection to only a single access point (AP), while L\&S demand simultaneous line-of-sight propagation paths to several APs, which serve as location and orientation anchors. Hence, a smart city deployment optimized for communication will be insufficient to meet stringent L\&S requirements. 
In this article, we argue that the emerging technologies of reconfigurable intelligent surfaces (RISs) and sidelink communications constitute the key to providing ubiquitous coverage for L\&S in smart cities with low-cost and energy-efficient technical solutions. 
To this end, we propose and evaluate AP-coordinated and self-coordinated RIS-enabled L\&S architectures and detail three groups of application scenarios, relying on low-complexity beacons, cooperative localization, and full-duplex transceivers. A list of practical issues and consequent open research challenges of the proposed L\&S systems is also provided. 
\end{abstract}



\section{Introduction}
The emerging concept of smart cities aims to improve accessibility to public services, advance digitization of the urban environment, and monitor various human-oriented processes as well as assets, by harmonizing diverse digital technologies at a city level~\cite{kisseleff2020reconfigurable}. This broad concept integrates various independent applications to bring improvements both at the societal level (such as smart homes, smart transportation, supply chains, and environment monitoring), and at the individual level (such as indoor navigation and extended reality (XR)). To realize the concept of smart cities, reliable, low-latency, and high-speed communication systems (to support information exchange and management among interconnected devices), as well as accurate \ac{las} (to support communication and provide situation-awareness services), are of great importance. In this article, we use the term \textit{localization} to indicate the \textit{position} (and possibly \textit{orientation}) estimation of a target \ac{ue}, and the term \textit{sensing} to specify the \textit{position estimation} of passive objects (i.e., objects without networking infrastructure or non-cooperating ones).

By exploiting the large antenna array sizes and wide bandwidth of millimeter-wave/THz systems, recent research activities in industry and academia on \ac{islac} are growing. \ac{islac} is able to utilize communication infrastructures and signals to enable synergies with \ac{las} for diverse applications. To this end, several standardization efforts and 3GPP activities have been recently studied, such as the definition of new radio positioning requirements, evaluation methodologies, and techniques (dependent on radio access technology and not), in TR 38.855~\cite{tr38855} as well as the development of Wi-Fi sensing technology (in both sub-7 GHz and mmWave spectrum), in the IEEE 802.11bf standard~\cite{chen2022wi}.



While high angular and delay resolution (due to large arrays and wide signal bandwidths)
facilitate \ac{las} tasks, signal coverage is one of the major challenges, especially for high-frequency systems which suffer from high path loss and increased blockage probability. 
{Although communication is possible with a link to a single \ac{ap} (e.g., a gNB macro base station), localization functions necessitate access to multiple \acp{ap} and sensing requires to combat the attenuation of \ac{tx}-object-\ac{rx} links, hence, L\&S service coverage can be quite limited}.
Reflective \acp{ris}, {which have been recently recognized as a promising technology for wireless communications in smart cities~\cite{kisseleff2020reconfigurable}, have a huge potential to improve L\&S performance, and even, enable L\&S services in various scenarios~\cite{wymeersch2020radio}.} 
However, since RISs are incapable of generating their own signals and only modify the analog waveforms impinging on them, separate signal generation sources are needed. 

{In 5G NR positioning~\cite{tr38855}, the downlink (from \ac{ap} to UE) positioning reference signals and uplink (from UE to \ac{ap}) sounding reference signals can serve as potential signal sources for RIS-aided L\&S systems. However, the deployment of power-hungry APs with full communication capabilities is costly, and UEs usually have limited power to provide sufficient signal strength.} To fulfill the ubiquitous \ac{las} requirements in out-of-coverage areas or partially-covered ones (e.g., indoor UEs and vehicles in tunnels), sidelink communication {(which refers to direct communication between terminal nodes or UEs without data going through the APs)} via the PC5 interface can be particularly useful~\cite{garcia2021tutorial, tr38845}. 
{This communication can alleviate high signal power loss (e.g., when the AP-UE link is of large distance) through denser deployment of low complexity beacons that only broadcast or receive L\&S signals. In addition, cooperative localization between several \acp{ue} can be employed to enable or enhance \ac{las}.}
Furthermore, when a \ac{ue} is equipped with a {full-duplex} transceiver \cite{FD_MIMO_arch} (a technology that is widely discussed in \ac{islac} for sensing purposes), it can both simultaneously transmit and receive the signal to localize itself via a single \ac{ris} anchor. {By combining sidelink communication and RISs, efficient AP-free L\&S services are feasible.}

\begin{figure*}[t]
\centering
\centerline{\includegraphics[width=0.7\linewidth]{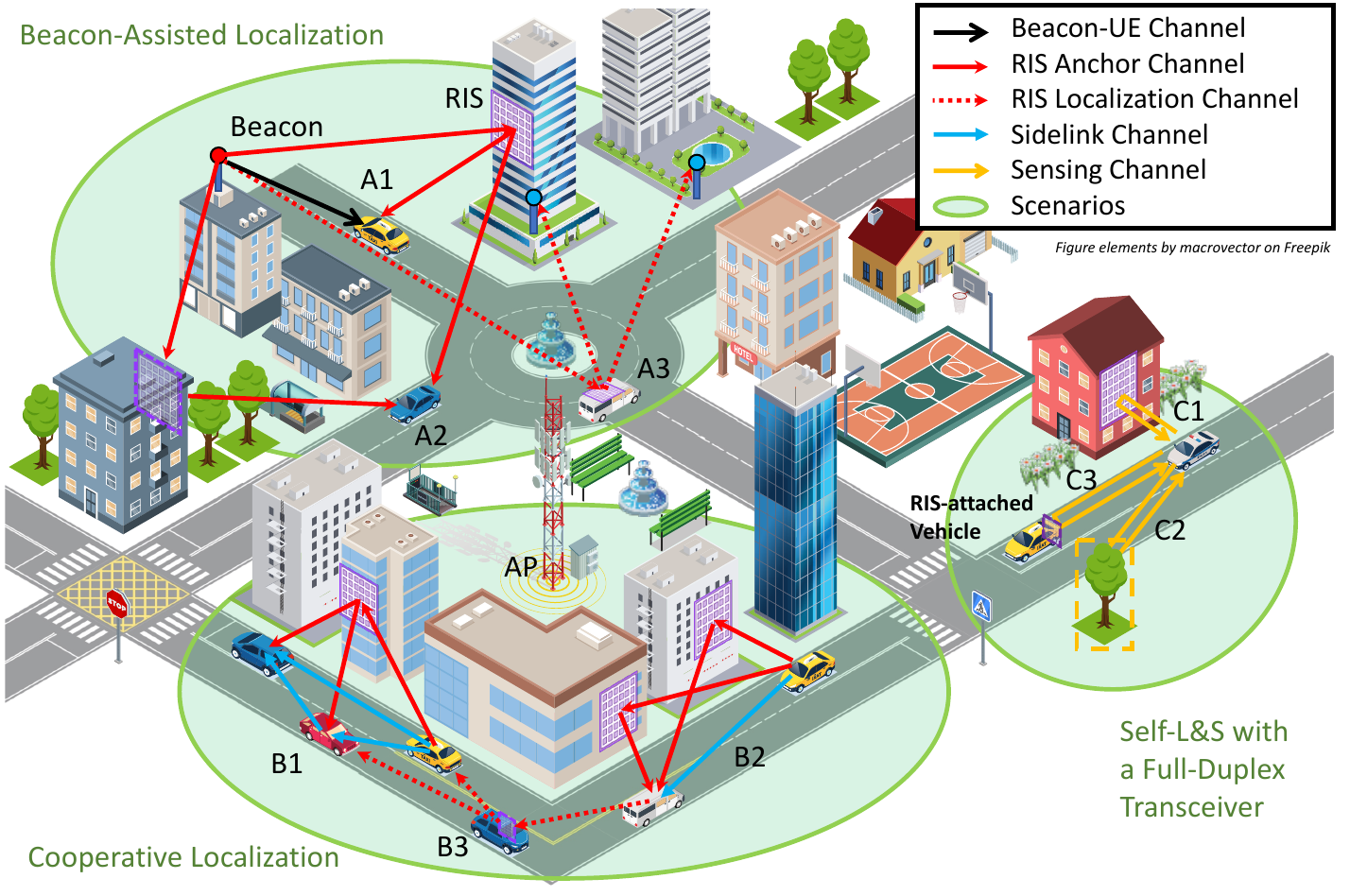}}
\caption{RIS-enabled seamless \ac{las} scenarios in smart cities: a) beacon-assisted localization, b) cooperative localization, and c) self-\ac{las} with a full-duplex transceiver. {The signals received from a RIS provide extra distance (via the path delay) and directional information (via the AOA, AOD, or spatial frequency), which enable or improve \ac{las}.}
} \vspace{-5mm}
\label{fig-1-illustration}
\end{figure*}

In this article, we argue that the RIS technology, in conjunction with sidelink communications, can provide \textit{seamless} \ac{las} {services to partially-covered and out-of-coverage areas, and thus,} 
speed up the intelligent transformation of cities into smart environments. {The contributions of this work are as follows:} {\textit{i)} AP-coordinated and AP-free \ac{las} architectures and protocols are proposed; \textit{ii)} representative RIS-enabled \ac{las} scenarios (i.e., \textit{beacon-assisted localization}, \textit{cooperative localization}, and \textit{self-\ac{las} with a full-duplex transceiver} as illustrated in Fig.~\ref{fig-1-illustration}) are discussed; and \textit{iii)} open challenges with the proposed RIS-enabled \ac{las} systems in smart cities are presented, together with potential directions for future research.}

\section{\ac{las} Architectures and Protocols}
\label{sec:architecture_description}
In this section, we describe the different entity types of the proposed \ac{las} system for smart cities, relying on \acp{ris} and sidelink communications, as well as its enabling architectures, depending on whether an \ac{ap} is present for \ac{las} coordination. 

\subsection{Entity Types and Overall Architecture}
\begin{figure}[t]
  \centering
\centerline{\includegraphics[width=0.99\linewidth]{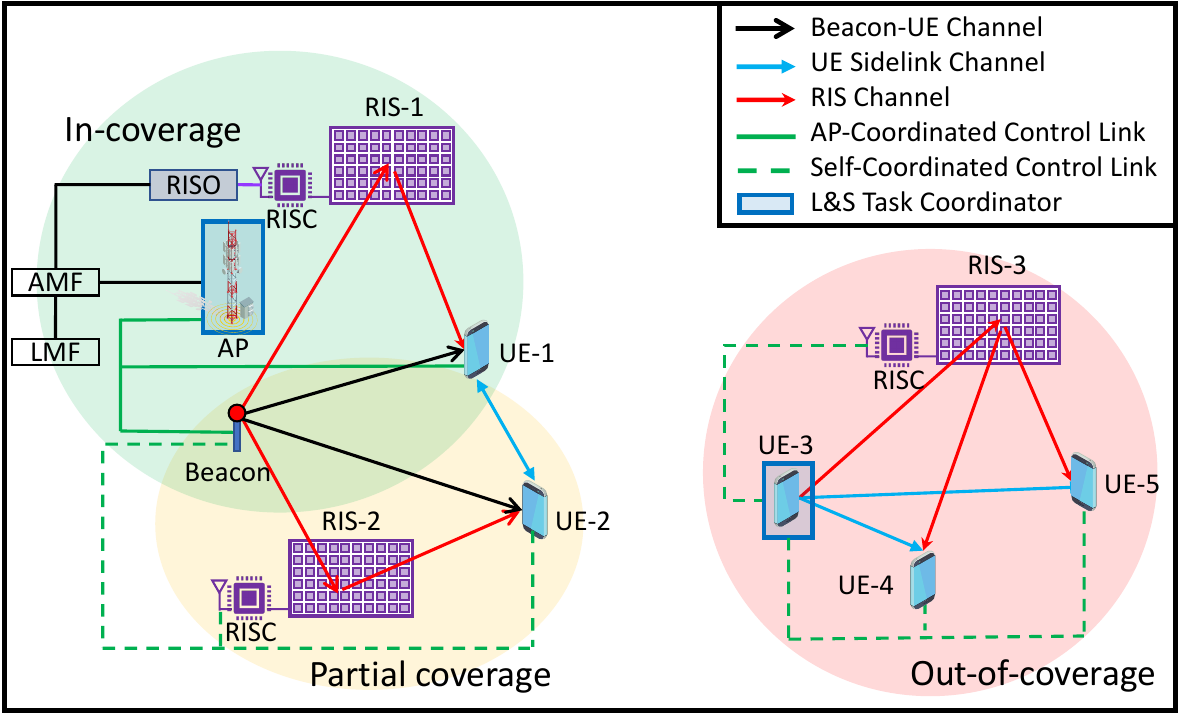}}
\caption{UEs performing \ac{las} in different coverage areas. The AP-coordinated architecture can be used for UEs located in in-coverage or partial-coverage (sidelink communications required) areas. When the UEs and all the \ac{las} devices cannot access any surrounding APs (i.e., lying in out-of-coverage areas), the self-coordinated architecture is the only option for \ac{las} services.}
\vspace{-5mm}
\label{fig-2-architecture}
\end{figure}

In the proposed RIS-enabled \ac{las} system, there are several types of entities: \acp{ap}, beacons, \acp{ris}, and \acp{ue}, as shown in Fig.~\ref{fig-2-architecture}. The \acp{ap} provide cellular services to the devices in coverage. A beacon could be a \ac{rsu} that is capable of sending and receiving \ac{las} reference signals via sidelink communications. In this way, the explicit involvement of expensive and power-hungry \acp{ap} with full communication protocols is not needed for \ac{las} purposes. 
RISs serve as reference anchors to assist ISLAC tasks, {each being controlled by a local \ac{risc}. This controller is able to switch RIS profiles autonomously or configured by other devices, for example, a UE via sidelink communication or other RISCs via a \ac{riso}~\cite{RISE6G_D51}.} Finally, UEs may have different hardware capabilities (e.g., single/multiple antennas, half-/full-duplex) that play different roles in \ac{las} (e.g., a target UE to be localized, an assistant UE with known or measured location to assist \ac{las}, or a server/coordinator in performing \ac{las} tasks). Although all the devices contributing to the targeted tasks will have sidelink communication capabilities, beacons usually have fewer power constraints than UEs, while {RISCs are expected to configure phases} in low-power mode without sending or processing \ac{las} reference signals~\cite{strinati2021reconfigurable}.
We further consider that \acp{ris} coordinate their reflective beamforming, either using time division or phase profile codes in the time domain, {which could be predefined or configured adaptively (e.g., optimized for L\&S performance using machine learning algorithms)}.

For the scenarios considered in this article, we propose two different architectures: one based on AP coordination and the other on self-coordination. The former architecture works for UEs (and other \ac{las}-related devices) inside a coverage area or in partially-covered areas (where sidelink is available), requiring a specific \ac{ap} to serve as a \ac{las} coordinator. {The \ac{las} tasks are performed by the interaction between the coordinator and the \ac{lmf} (e.g., via the NR positioning protocol A (NRPPa)~\cite{tr38455}), which controls and coordinates all involved devices via the \ac{amf}, as shown in Fig.~\ref{fig-2-architecture}.} 
The self-coordinated architecture relies explicitly on sidelink communications, being particularly suitable for UEs in the out-of-coverage of \acp{ap}, or for UEs connected to \acp{ap} that cannot meet the latency and spatial resolution requirements (e.g., legacy 3G/4G \acp{ap}). {In this case, the \ac{las} are performed at the selected task coordinator (e.g., a UE).} In both architectures, \ac{las} signals can be generated by a beacon, an assistant UE, or a target UE, depending on the network topologies and the specific application scenario. {Although L\&S tasks require dedicated algorithms, they can be performed using the following architectures and protocols.}

\subsection{AP-Coordinated Architecture}
\label{sec:architecture_with_aps}
This architecture relies on one or several APs to allocate the available radio resources and ensure timing among the connected devices, which are the UEs, the RIS controllers, and dedicated beacon nodes. This network management scheme is similar to the mode-1 in sidelink communications~\cite{garcia2021tutorial}, and the \ac{las} protocols can be summarized into the following $6$ steps:
\begin{enumerate}
\item The target UE triggers an \ac{las} request to the AP (which is selected as the task coordinator).
\item The coordinator exchanges related location information {(e.g., surrounding beacons and RISs that are registered in the system), determines \ac{las} configurations (e.g., communication mode and RISC setting)}, as well as selects and notifies nearby beacons, UEs, and RISs that are involved in the \ac{las} task. 
\label{test_step2}
\item The coordinator allocates time-frequency resources for \ac{las} to all involved devices, and {triggers the \acp{risc} to configure the phase profiles of the RIS elements} for the whole duration of the estimation process.
\item The beacons and/or UEs transmit \ac{las} reference signals, which are reflected by the involved RIS(s) and received by the target UE. 
\item The collected measurements are used by the target UE for localization and/or sensing tasks. Alternatively, this computation can be offloaded at the L\&S server (e.g., the coordinating AP or another UE with high computational power). {For passive reflective RISs, the RISC does not have access to the impinging radio signal and thus cannot perform any estimation.} 
\item The task coordinator is updated with the estimated \ac{las} results; this optional step can serve as prior information for future use.
\label{test_step6}
\end{enumerate}

For the partial-coverage scenarios, the out-of-coverage UEs need to establish sidelink communications with devices that are covered by \acp{ap}. Then, the \ac{las} tasks can be performed similarly to the aforementioned steps. 


\subsection{Self-Coordinated Architecture}
An AP-free architecture is required for cases where the devices involved in \ac{las} tasks are located in out-of-coverage areas. Similar to the mode-2 sidelink communications~\cite{garcia2021tutorial}, \ac{las} tasks can be autonomously realized by selecting a specific device as the localization coordinator, as follows: 
\begin{enumerate}
    \item The target UE discovers nearby devices (e.g., beacons, RISs, and other UEs) {via sidelink communication} and obtains their location information (if available).
    \item Based on the discovered neighbors, the target UE determines a \ac{las} task coordinator (could be itself) and notifies it of the \ac{las} configurations.
    \item The target UE triggers a \ac{las} request to the coordinator and performs the same actions as with the AP-coordinated architecture (i.e., steps \ref{test_step2})--\ref{test_step6})).
\end{enumerate}

{
In the following section, we describe the selected \ac{las} scenarios with RISs and sidelink communications illustrated in Fig.~\ref{fig-1-illustration}. Scenarios A and C were recently presented in~\cite{keykhosravi2022ris, Hyowon_RIS-SLAM_TWC2022}; however, the more general cases of 3D positioning, orientation estimation, and mobility, as well as the scenarios B1-B3 have not been reported, to the best of our knowledge, in the existing open technical literature.}

\section{RIS-Enabled \ac{las} Scenarios}
In this section, we will present three representative RIS-enabled \ac{las} scenarios for smart city applications relying on the aforementioned architectures and protocols. 


\subsection{Beacon-Assisted Localization}
\label{sec:1_single_antenna_beacon}

With low-complexity beacons, high flexibility in the installation and deployment of \ac{las} systems is feasible. A typical use case could be \textit{a train station} with multiple low-cost beacons broadcasting \ac{las} reference signals to \acp{ue} to navigate indoors, via the support of \acp{ris}.
%
%
In this category, 
we consider UE localization (with the aid of one or more RISs) and RIS localization (with the aid of several beacons). 

\subsubsection*{{\rm A1)} \textbf{Single-RIS-Enabled UE Localization}}
{In this scenario, the RIS and beacon states are assumed to be known so that an extra delay and the~\ac{aod} of the RIS path can be estimated from the beacon's \ac{las} reference signals.}
The target UE can then be localized by the intersection of a hyperbola (i.e., \ac{tdoa} of the LOS and RIS paths) and the line in the direction of the~\ac{aod} at the RIS. {Strong multipath will affect the performance~\cite{keykhosravi2022ris}; however, with proper RIS profile design, controllable RIS channels can be separated from uncontrolled multipath channels, leading to better performance.} 
This is the basic RIS-enabled localization scenario that only requires a single low-complexity beacon.

\subsubsection*{{\rm A2)}  \textbf{Multi-RIS-Enabled UE Localization}}
\begin{figure}[!t]
  \centering
\centerline{\includegraphics[width=1.15\linewidth]{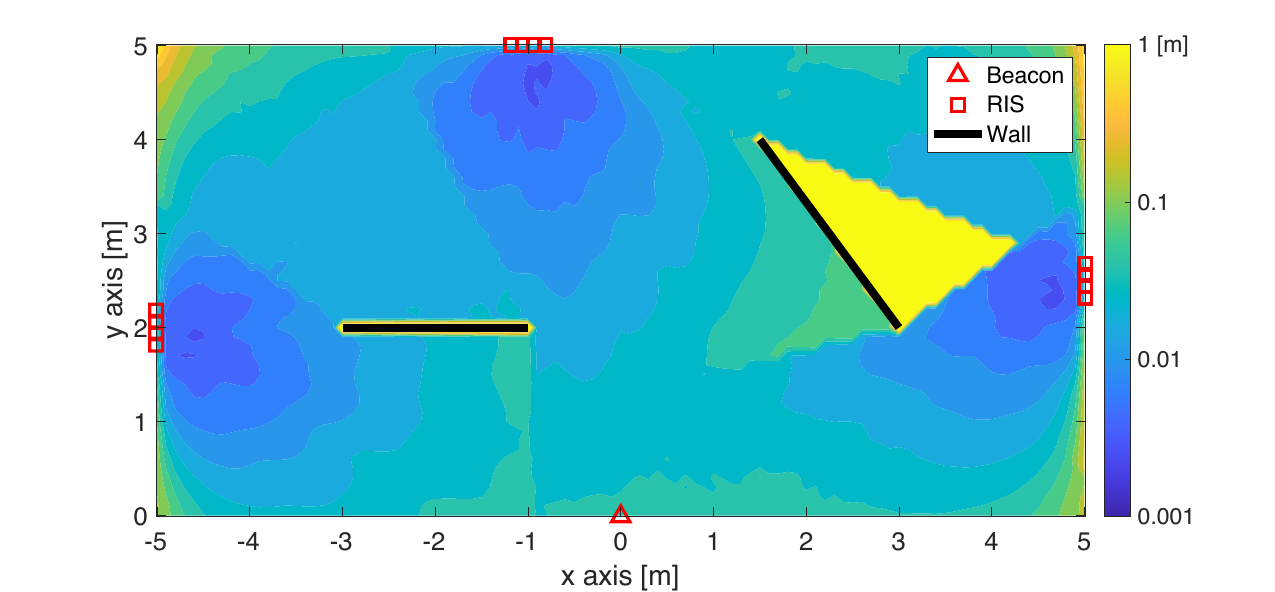}}
\caption{Scenarios A1 and A2: \ac{peb} (in meters) with different UE positions in a multi-RIS-aided {SISO} localization scenario {at $28$~GHz with $10\times 10$ RISs}. The target UE can be localized with a single \ac{ris} and a beacon-UE LOS path, or with at least $2$ RISs under LOS blockage conditions {(see~\cite{keykhosravi2022ris} for more details).}
} \vspace{-5mm}
\label{fig-3}
\end{figure}

If multiple RISs are simultaneously available, the requirements for LOS and delay estimations are unnecessary. The \acp{aod} from different RISs can be estimated and used to localize the UE by intersecting the \ac{aod} lines. Fig.~\ref{fig-3} shows the \ac{peb} of the target UE with different positions inside a $5\times 10$ m$^2$ area. As shown, with multiple RISs, the UE is localizable even under blockage of the LOS path between the beacon and the UE. However, the localization tasks cannot be performed when only one anchor (beacon or RIS) is visible to the UE (see the yellow triangular area around the point $[3,3]$ m).

\subsubsection*{{\rm A3)} \textbf{RIS Localization via Multi-Static Sensing}}
In a scenario where passive \acp{ue} or objects are coated with RISs, the localization (or sensing, depending on scenarios) can be performed semi-passively with only a small amount of energy needed for localization coordination and RIS phase profile control. Such localization tasks can estimate the positions (and orientations) of RIS-coated objects by using
several beacons with known positions {and the estimated delays and spatial frequencies (rather than the AODs used in A1 and A2)}. Note that the geometrical constraints can largely reduce the difficulties in these scenarios. For example, the orientation of a RIS can be assumed as 1D (e.g., vehicle heading direction). In addition, the adoption of antenna arrays at the beacons can further simplify the RIS localization problem.



\subsection{Cooperative Localization}
\label{sec:2_cooperative_sidelink}
Sidelink communications, or the vehicular version known as \ac{v2x} communications,
has been introduced in the millimeter-wave band for information exchange between vehicles, opening the road for numerous use cases, such as platooning, collision avoidance and autonomous driving \cite{garcia2021tutorial}. The combination of \acp{ris} and sidelink is expected to provide low-latency and high-reliability communications~\cite{gu2022intelligent}, but also assist \ac{las} in a cooperative manner~\cite{ko2021v2x}. We will focus on the latter case, where the absolute location can be estimated using \ac{ris} anchors (i.e., with known position and orientation information), without any \acp{ap} or beacons involved. A typical scenario could be \textit{cooperative vehicular networks} in urban areas with severe \ac{ap} and GPS signal blockages. We consider single- and multi-\ac{ris}-involved localization scenarios, where single-antenna \acp{ue} cooperate to estimate their positions via sidelink signals. We also consider a more general scenario where \ac{ris}-coated objects are involved, resulting in cooperative \ac{ris} localization.

\subsubsection*{{\rm B1)} \textbf{Single-RIS-Enabled Cooperative Localization}}
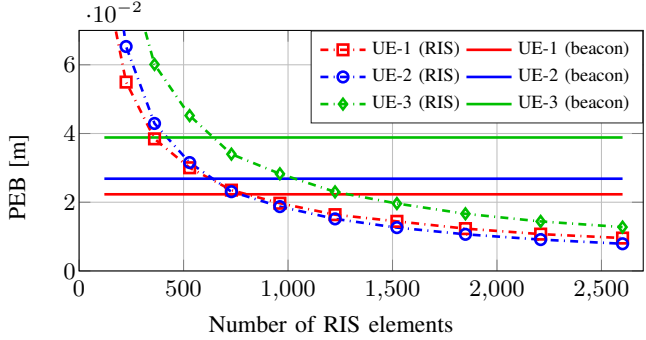
\begin{figure}[!t]
  \centering
\centerline{
%
\definecolor{mycolor1}{rgb}{0.46667,0.67451,0.18824}%
\definecolor{mycolor2}{rgb}{0, 0.75, 0}%

\begin{tikzpicture}

\begin{axis}[%
width=75mm,
height=32mm,
at={(0mm,0mm)},
scale only axis,
xmin=000,
xmax=2700,
ymin=0,
ymax=0.07,
yticklabel style = {font=\small,xshift=0.5ex},
xticklabel style = {font=\small,yshift=0ex},
axis background/.style={fill=white},
xmajorgrids,
ymajorgrids,
legend columns=2, 
legend style={font=\scriptsize, at={(1.0, 1.0)}, anchor=north east, legend cell align=left, align=left, draw=white!15!black}
]

\addplot [color=red, dashdotted, line width=1.0pt, mark=square, mark options={solid, red}]
  table[row sep=crcr]{%
121	0.08644530467154\\
225	0.0549642746013048\\
361	0.038444287455133\\
529	0.0300336128951666\\
729	0.0235150693327994\\
961	0.0197059857647924\\
1225	0.0163776884112299\\
1521	0.0144381640769358\\
1849	0.0123178501354005\\
2209	0.0107204061335489\\
2601	0.00957158998495567\\
};
\addlegendentry{UE-1 (RIS)}

\addplot [color=red, line width=1.0pt]
  table[row sep=crcr]{%
121	0.0223079327331444\\
225	0.0223079327331444\\
361	0.0223079327331444\\
529	0.0223079327331444\\
729	0.0223079327331444\\
961	0.0223079327331444\\
1225	0.0223079327331444\\
1521	0.0223079327331444\\
1849	0.0223079327331444\\
2209	0.0223079327331444\\
2601	0.0223079327331444\\
};
\addlegendentry{UE-1 (beacon)}

\addplot [color=blue, dashdotted, line width=1.0pt, mark=o, mark options={solid, blue}]
  table[row sep=crcr]{%
121	0.110113833888614\\
225	0.0653194834295803\\
361	0.0429235560786993\\
529	0.0315441840931914\\
729	0.0230674700390041\\
961	0.0187271952545166\\
1225	0.0151517942785312\\
1521	0.012615173056805\\
1849	0.0106633949153502\\
2209	0.00912977619810484\\
2601	0.00791131978849029\\
};
\addlegendentry{UE-2 (RIS)}

\addplot [color=blue, line width=1.0pt]
  table[row sep=crcr]{%
121	0.0268484300671912\\
225	0.0268484300671912\\
361	0.0268484300671912\\
529	0.0268484300671912\\
729	0.0268484300671912\\
961	0.0268484300671912\\
1225	0.0268484300671912\\
1521	0.0268484300671912\\
1849	0.0268484300671912\\
2209	0.0268484300671912\\
2601	0.0268484300671912\\
};
\addlegendentry{UE-2 (beacon)}

\addplot [color=mycolor2, dashdotted, line width=1.0pt, mark=diamond, mark options={solid, mycolor2}]
  table[row sep=crcr]{%
121	0.148081581958508\\
225	0.0898851703750865\\
361	0.0600993779960515\\
529	0.0452182904872591\\
729	0.033995437451595\\
961	0.0282788949215034\\
1225	0.0230148571487983\\
1521	0.0196225603777511\\
1849	0.016635682808083\\
2209	0.0144125630537031\\
2601	0.0127675584341733\\
};
\addlegendentry{UE-3 (RIS)}

\addplot [color=mycolor2, line width=1.0pt]
  table[row sep=crcr]{%
121	0.0388413293373138\\
225	0.0388413293373138\\
361	0.0388413293373138\\
529	0.0388413293373138\\
729	0.0388413293373138\\
961	0.0388413293373138\\
1225	0.0388413293373138\\
1521	0.0388413293373138\\
1849	0.0388413293373138\\
2209	0.0388413293373138\\
2601	0.0388413293373138\\
};
\addlegendentry{UE-3 (beacon)}


\end{axis}

\node[rotate=0,fill=white] (BOC6) at (3.35cm,-.7cm){\small Number of RIS elements};
\node[rotate=90] at (-8mm,12.5mm){\small PEB [m]};

\end{tikzpicture}%
}
\caption{\acp{peb} of \ac{ris}-enabled vs. beacon-aided cooperative localization for different \ac{ris} sizes and a fixed random phase profile setup. It is shown that, with a sufficient number of \ac{ris} elements, an active anchor {($3\times 3$ beacon array with $10\ \text{dBm}$ transmit power and $5$ transmissions)} can be replaced with a passive one (\ac{ris}) without performance degradation.}
\label{fig-4}
\end{figure}

Consider a scenario with several \acp{ue} and one \ac{ris} anchor, where the \acp{ue} wish to estimate their positions. We assume that each \ac{ue} can send sidelink signals (i.e., being the \ac{tx}) to other \acp{ue}, which arrive at the \ac{rx} \acp{ue} via two paths (i.e., the \ac{ue}-\ac{ue} and \ac{ue}-\ac{ris}-\ac{ue} paths). By proper control of the \ac{ris} elements, those two paths can be separated.
{Then, the delays for different paths and the spatial frequency information (similar to A3) related to each RIS can be estimated via the received symbols at the RX. Finally, those collected measurements can be utilized to estimate the locations of the TX and RX.}
This scenario requires at least three \acp{ue} to cooperate and render their locations feasibly without ambiguities. Fig.~\ref{fig-4} compares the \acp{peb} for three \acp{ue} in a \ac{ris}-enabled versus beacon-aided 3D cooperative localization scenario as a function of the number of \ac{ris} elements.

\subsubsection*{{\rm B2)} \textbf{Multi-RIS-Enabled One-Way Sidelink Localization}} In the scenario with at least two \acp{ris}, one-way sidelink communications is sufficient to localize both the TX and RX \acp{ue}. With two \acp{ris}, three delay measurements can be obtained between the TX and RX via the \ac{los} and the two \ac{ris} paths. However, that would require an optimal joint design of the reflection elements at both \acp{ris} to be able to separate the paths at the RX. In addition, once the \ac{ris} paths are separated, we can also estimate the spatial frequency information at each \ac{ris}. Thus, those collected measurements can be utilized to estimate the locations of the TX and RX.

\subsubsection*{{\rm B3)} \textbf{Cooperative RIS Localization}}
Consider a more general scenario where one \ac{ris} (or several) with a known state is used to localize multiple \acp{ue} (with sidelink capabilities) and objects (coated with a \ac{ris}). 
This scenario is challenging due to the high complexity of the network and a large number of unknowns. However, with a proper design of all the involved \ac{ris} profiles and the transmission protocol, this problem can be decomposed into a cooperative localization (see B1) and a \ac{ris} localization (see A3) problems. Similar to B1, at least several \acp{ue} (depending on scenarios) need to take the role of the TX, and transmit sidelink signals to the other \acp{ue} via the direct and indirect paths. Once the \acp{ue} are localized, the \ac{ris} localization task can be solved similarly to A3, and the estimation results can be refined by processing all the available information. 





\subsection{Self-\ac{las} with a Full-Duplex Transceiver}
\label{sec:3_self_positioning_full_dulplex_radar}



When a \ac{ue} is equipped with a full-duplex transceiver (like radar)~\cite{FD_MIMO_arch}, the multi-\ac{ris} setup and cooperation between \acp{ue} are unnecessary. Instead, this \ac{ue} can perform self-positioning with a single \ac{ris} and use the multipath components to map the environment over time; this process is known as monostatic \ac{slam}.  
It is noted that \ac{slam} is not limited to full-duplex \acp{ue}, and bistatic \ac{slam} can also be performed in use cases A1-A3 and B1-B3. 
We next present three beacon-free \ac{las} scenarios with a full-duplex \ac{ue}. 

\subsubsection*{{\rm C1)} \textbf{RIS-Enabled Self-Localization}}
\label{application:C1}
\begin{figure}[!t]
\begin{minipage}[b]{0.98\linewidth}
  \centering
\centerline{\includegraphics[width=0.9\linewidth]{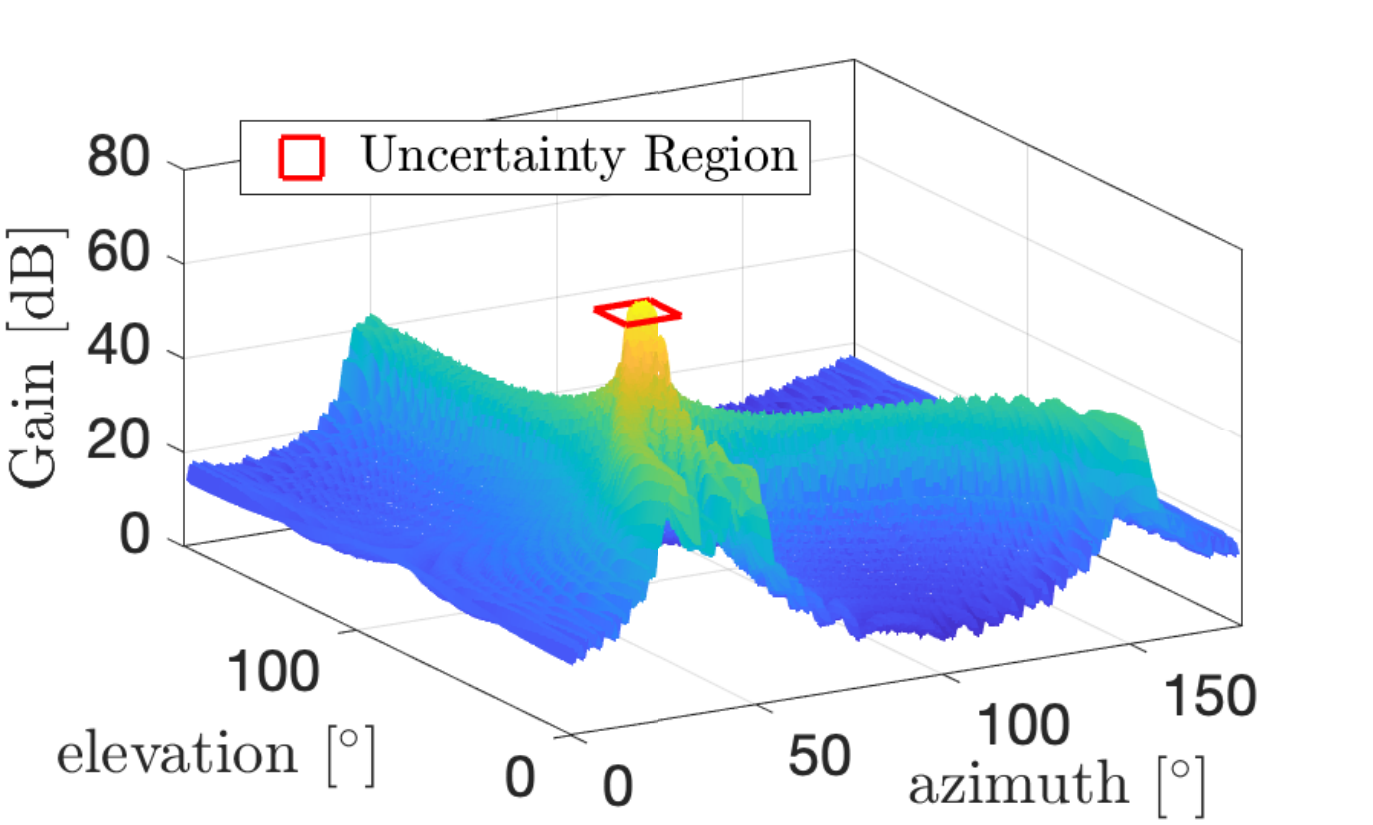}}
\centerline{\small{(a) Directional Profiles}} \medskip
\label{Fig:directional}
\end{minipage}
\begin{minipage}[b]{0.98\linewidth}
  \centering
\centerline{\includegraphics[width=0.9\linewidth]{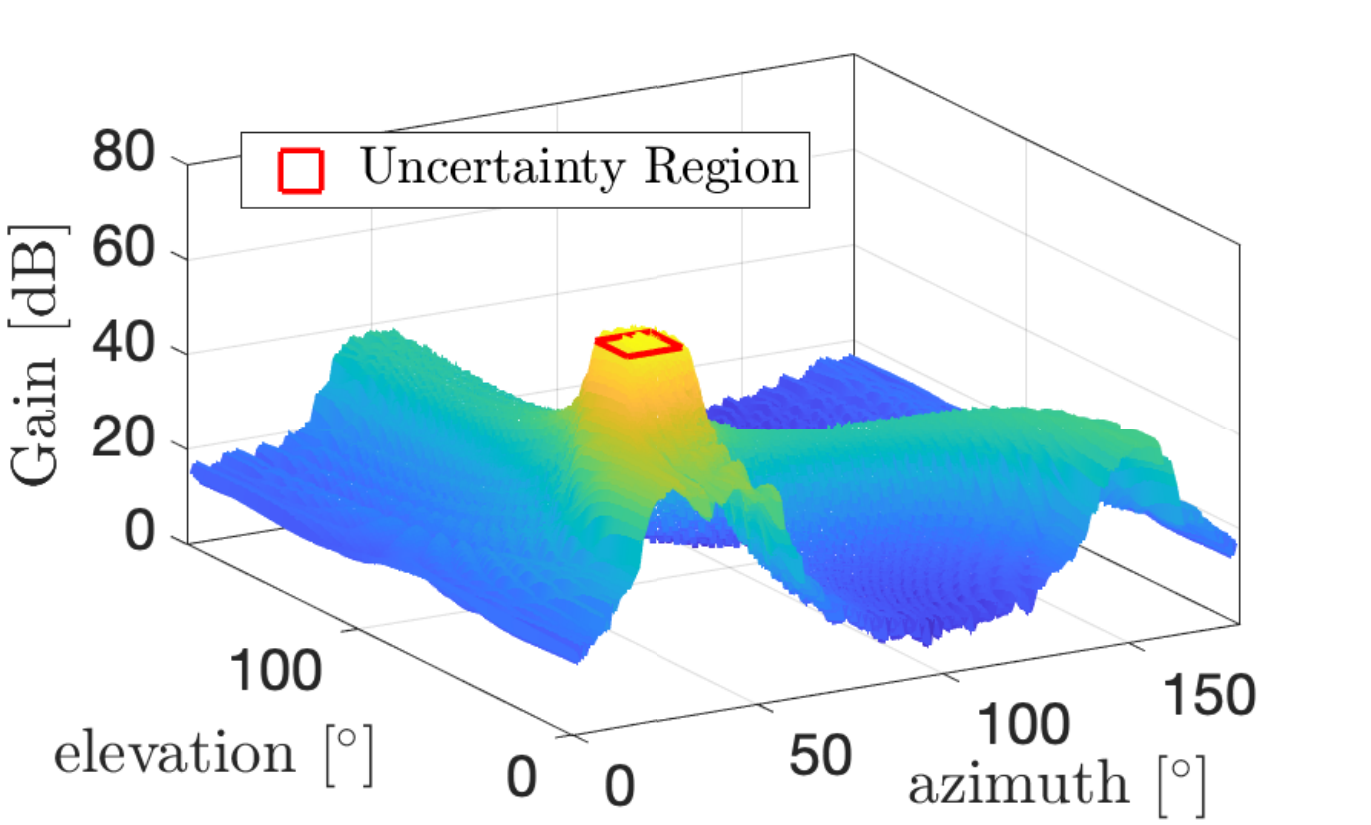}}
\centerline{\small{(b) Optimized Profiles}} \medskip
\label{Fig:optimized}
\end{minipage}
\caption{The reflective beamforming gain (in dB) with a RIS using (a) a directional profile, and (b) an optimized profile via~\cite{Hyowon_RIS-SLAM_TWC2022} {($50\times 50$ RIS arrays, 20 transmissions, $20^\circ$ and $16^\circ$ angular uncertainty in azimuth and elevation)}. The red squares represent the UE angular uncertainty region that needs to be covered. 
}
\label{fig-5}
\end{figure}

Consider a system with a single-antenna UE and a RIS, where the UE transmits \ac{las} reference signals and receives their back-scattered versions, i.e., the UE-RIS-UE (controlled path) and UE-landmark-UE (uncontrolled path) signals. 
One option for the RIS phase profiles is to consider directional reflective beams, which can be efficiently designed when the \ac{ue} position uncertainty (even under mobility cases) is available~\cite{Hyowon_RIS-SLAM_TWC2022}.
The delay and angle information at the RIS of the UE-RIS-UE channel can be estimated for this scenario, and then used to localize the UE. 
In Fig.~\ref{fig-5}, beampatterns at the RIS with two different phase profiles are illustrated, focusing on the \ac{ue} uncertainty region. As demonstrated, the optimized phase profiles of~\cite{Hyowon_RIS-SLAM_TWC2022} can provide sufficient beamforming gain, compared with directional phase profiles. This gain can offer improved \ac{las} performance. 



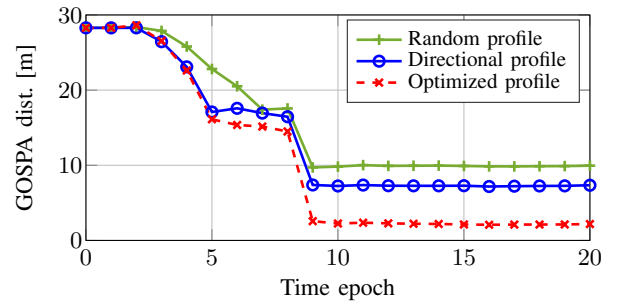
\begin{figure}[!t]
  \centering
\centerline{
%
%
\definecolor{mycolor1}{rgb}{0.46667,0.67451,0.18824}%

\begin{tikzpicture}

\begin{axis}[%
width=67mm,
height=30mm,
at={(0mm,0mm)},
scale only axis,
xmin=0,
xmax=20,
ymin=0,
ymax=30,
yticklabel style = {font=\small,xshift=0.5ex},
xticklabel style = {font=\small,yshift=0ex},
axis background/.style={fill=white},
axis background/.style={fill=white},
xmajorgrids,
ymajorgrids,
legend style={legend cell align=left, align=left, draw=white!15!black,style={row sep=-0.1cm}}
]

\addplot [color=mycolor1, line width=1.0pt, mark=+, mark options={solid, mycolor1}]
  table[row sep=crcr]{%
0	28.2842712474619\\
1	28.2842712474619\\
2	28.3858053003673\\
3	27.879239682682\\
4	25.7959264744949\\
5	22.8111184495208\\
6	20.5079611610609\\
7	17.4026892336108\\
8	17.5634695858882\\
9	9.71473638492555\\
10	9.82123594197519\\
11	10.0183426362617\\
12	9.92013903136244\\
13	9.93814222424261\\
14	9.95991556484288\\
15	9.90792379014649\\
16	9.85632048683771\\
17	9.84959106699165\\
18	9.87305906284139\\
19	9.89073110778893\\
20	9.9461290514263\\
};
\addlegendentry{\footnotesize{Random profile}}

\addplot [color=blue, line width=1.0pt, mark=o, mark options={solid, blue}]
 table[row sep=crcr]{%
0	28.2842712474619\\
1	28.2842712474619\\
2	28.2842712474619\\
3	26.4438778236754\\
4	23.0926646637506\\
5	17.0884236713215\\
6	17.5789057898801\\
7	16.9414335631823\\
8	16.4580821142444\\
9	7.38155287282108\\
10	7.24696911316592\\
11	7.36670489680816\\
12	7.28139085816603\\
13	7.26627758965934\\
14	7.26684432284794\\
15	7.28074534395769\\
16	7.17515038724541\\
17	7.21791131122367\\
18	7.24670596218139\\
19	7.26098555999293\\
20	7.34366448763454\\
};
\addlegendentry{\footnotesize Directional profile}

\addplot [color=red, dashed, line width=1.0pt, mark=x, mark options={solid, red}]
  table[row sep=crcr]{%
0	28.2842712474619\\
1	28.2842712474619\\
2	28.6181217828841\\
3	26.5477857079029\\
4	22.6092719824074\\
5	16.1208158425622\\
6	15.3624605680139\\
7	15.1517620710929\\
8	14.4965170063703\\
9	2.56344729062067\\
10	2.25300786924345\\
11	2.35693483527853\\
12	2.27175060688199\\
13	2.22079421742703\\
14	2.18946381478161\\
15	2.12422577756531\\
16	2.08328806530829\\
17	2.10797495903972\\
18	2.11501354148465\\
19	2.12568478631453\\
20	2.18055114305785\\
};
\addlegendentry{\footnotesize Optimized profile}
\end{axis}

\node[rotate=0,fill=white] (BOC6) at (3.35cm,-.65cm){\small Time epoch};
\node[rotate=90] at (-8mm,15mm){\small GOSPA dist. [m]};
\end{tikzpicture}%
}
\caption{{The evaluation of RIS-enabled SLAM performance, in terms of generalized optimal subpattern assignment distance (GOSPA), with a $20\times20$ RIS with different phase profiles (random, directional, and optimized via \cite{Hyowon_RIS-SLAM_TWC2022}), using $20$ transmissions and a $4\times4$ UE array in an environment with $4$ visible landmarks).}
}
\label{fig-6}
\end{figure}

\subsubsection*{{\rm C2)} \textbf{RIS-Enabled SLAM}} 
\label{application:C2}
If a UE is equipped with a full-duplex multiple-input multiple-output (MIMO) antenna array \cite{FD_MIMO_arch}, SLAM can be enabled. Similar to scenario C1, the signals from different paths can be resolvable with optimized \ac{ris} phase profiles and precoders/combiners. 
The following channel parameters can be estimated: \textit{i)} the signal propagation delay, the \ac{aod} at the RIS, and the \ac{aoa} at the UE for the UE-RIS-UE channel; as well as \textit{ii)} the delay and \ac{aoa} at the UE for each UE-landmark-UE channel. 
In addition to the position information obtained from the controlled path (as in scenario C1), {the parameters of the uncontrolled paths reaching the UE array are exploited to sense the propagation environment. Using state-of-the-art SLAM filtering, UE localization and radio mapping (i.e., sensing) performance can be improved.}
{The evaluation of the sensing performance with different phase profiles is illustrated in Fig.~\ref{fig-6}, showcasing the prominent role of the RIS reflection optimization.}

\subsubsection*{{\rm C3)} \textbf{RIS Localization with a Full-Duplex Array}}
Consider the more general scenario from C1 including one anchor RIS mounted on a wall, a UE equipped with a full-duplex MIMO transceiver, and several objects coated with RISs (e.g., mounted on the front and rear side of a vehicle). In addition to the signals reflected from the anchor RIS (as also in scenario C1), the UE also receives single-bounce reflected signals from the RISs  mounted on the objects. 
The time delay, \ac{aoa}, and the amplitude of the channel gain for each signal path can be estimated, which can be used for the localization of both itself and the RISs-coated objects. 
When multiple UEs are present and cooperate in the estimation process, the orientation of the RISs-coated objects can also be obtained. In a scenario without any anchors, this \ac{ris} localization can also help in estimating the relative locations of the active \ac{ue} and passive \acp{ue}. 




\section{Open Research Challenges}
With the assistance of low-complexity beacons, cooperative localization, and full-duplex radios, the \ac{las} coverage for smart city applications can be significantly extended. However, there exist several practical issues that need to be thoroughly investigated. In this section, we discuss the most critical challenges with the proposed RIS-enabled \ac{las} system and list possible directions for future research.

\subsection{Anchor Deployment Optimization}
The placement of the anchors (e.g., beacons and RISs) is critical to meet the \ac{las} \ac{kpi} requirements within a service area (e.g., error bounds lower than a certain threshold, as shown in Fig.~\ref{fig-3}). {In scenarios without any active anchors (e.g., B1-B3 and C1-C3), the L\&S design needs to take into account the state distributions and hardware capabilities of all involved devices.} The deployment involves both the position and orientation optimization of the anchors,  taking into account the blockage in the surrounding environment. 
RIS-aided SLAM can help in creating such an environment map, which can be supported by cooperative sidelink \acp{ue}. Heuristic optimization solutions can then be applied to finding optimal anchor sites.


\subsection{Resource Allocation and Coordination}
Resource allocation for \ac{las} tasks, including power and time-frequency allocation, beamforming design, and scheduling, must be carefully designed to ensure a favorable trade-off with conventional communication services.
Depending on the \ac{kpi} requirements of the applications that send \ac{las} service requests, new objectives that consider integrated \ac{las} and communications should be formulated and satisfied. 
An important part of resource allocation is RIS phase profile optimization and multiplexing~\cite{keykhosravi2022ris}. 
Broad RIS beams lead to coverage reduction, while narrow pencil beams are sensitive to misalignment. 
Hence, highly adaptive RIS profile designs are needed, relying, when possible, on prior UE and object state information. 
{When dealing with multiple RISs and UEs, resource allocation becomes more complex, and the increased coordination overhead needs to be accounted for.}
RIS multiplexing can be addressed by time multiplexing, temporal coding, and making use of high path loss for spatial reuse. 
%
%
The afore-described resource allocation problems can be tackled by a combination of traditional optimization-based methods (e.g., convex optimization) and learning-based methods (e.g., reinforcement learning).

\subsection{Estimation Algorithms}
From an algorithmic perspective, there are challenges related to channel parameter estimation, tracking in dynamic environments, and anchor calibration. 
{Different from most existing works that assume the channel between the RIS and the active signal source (e.g., an AP) to be known, the AOAs/AODs at the RISs are coupled in cooperative localization (scenarios B1 and B2) and RIS localization (e.g., scenarios A3, B3, and C3) tasks, requiring novel algorithms to process the estimated spatial frequencies.} More refined channel parameter estimation also requires accurate channel models and the RISs' impact on them, such as the near-field effect, beam squint effect, RIS element failures, and hardware impairments. 
Due to mobility, difficult conditions such as signal blockage, unresolvable signal paths, and severe path loss will affect \ac{las} performance. Multiple \acp{ris} can be involved to handle such blockages, offering coverage extension.
Sensing also suffers from inherent complications, such as
an unknown number of objects, unknown types of objects, unknown detection probabilities for signal paths, extended objects, and multi-bounce observations. Dedicated filters should be developed to address these complications and get integrated into the \ac{las} framework. Finally, in terms of anchor calibration, which is similar to RIS localization (as described in scenarios A3, B3, and C3), requiring a calibration agent that incorporates other sources of localization estimations (i.e., sensor fusion). 

\subsection{Understanding Anchor Hardware Alternatives}
There are also opportunities to improve \ac{las} coverage via variations of the hardware deployed at the beacons, RISs, and UEs.  On the beacon and UE sides, multi-panel arrays (i.e., 3D arrays) could be implemented for further coverage extension. On the RIS side, new types of RISs are emerging beyond almost passive reflective RISs~\cite{Tsinghua_RIS_Tutorial}. As previously mentioned, an active RIS can be used to boost the signal energy (i.e., change both the amplitude and phase of the incident signal) for improved coverage. A receiving RIS (also known as a hybrid RIS or a simultaneously reflecting and sensing RIS) can enable parameter estimation at the RIS side, offering extra degrees of freedom for the design of \ac{las} estimation approaches. Omni-directional RISs, intended to realize simultaneous reflection and refraction (i.e., $360^{\circ}$ coverage), enable simultaneous indoor and outdoor 3D localization.
A non-reciprocal RIS that integrates nonreciprocal phase shifters allows full-duplex communications, and
a delay-adjustable RIS is capable of adjusting the delays of signals reflected by different RIS elements, which contributes to the alleviation of the beam squint effect. All of these alternatives have implications on \ac{las} services and merit further study.

\subsection{Privacy, Security, and Social Acceptance Issues}
Cooperative L\&S require extensive information exchange of local measurements between devices, which may cause privacy issues. {For example, the RX in scenario B2 can estimate the position of the TX with a one-way pilot signal transmission, which may lead to a disclosure of private information.} In addition, 
different types of cyber attacks can reduce the \ac{las} service availability, or even provide an undetected erroneous location estimation, which is unacceptable for safety-critical applications.
Currently, several security management systems have been standardized (e.g., IEEE 1609.2), and security threats have been identified for sidelink communications. 
However, the discussions on \ac{las} task-related security issues are still at the initial stage, and potential threats need to be explored and eliminated.
A final aspect related to the widespread adoption of RISs lies in their social acceptance. RISs should be integrated in a way that they blend into the environment (ideally, be transparent). To this end, the benefits of RISs to improve safety and reduce electromagnetic emissions should be demonstrated and disseminated. 
{In summary, in addition to providing seamless L\&S services via RISs and sidelink communications, the issues of privacy, security, and social acceptance are extremely important deserving further research and regulation development.}
\section{Conclusion and Outlook}
\label{sec:opportunities_and_challenges}
The smart city paradigm constitutes the epitome of the widespread adoption of digital services for societal needs. It is envisioned to profit people and city-level businesses, offering efficient, safe, and comfortable living spaces as well as everyday-life smart-living applications. To achieve this overarching goal, seamless wireless communications among diverse devices and \ac{las} are of paramount importance, enabling information exchange, device localization, and mapping of the environment. In this article, we discussed the key to achieving low-cost and energy-efficient seamless \ac{las}, namely, reflective RISs in conjunction with sidelink communications. We presented AP-coordinated and AP-free system architectures and detailed three RIS-enabled \ac{las} scenarios, each including several use cases and most relying on sidelink communications. As became apparent, instead of using APs with full communication capabilities, low-complexity beacons and RISs can be widely-deployed to enable green \ac{las} smart city applications. In addition, when multiple UEs with sidelink communication capabilities can be connected in the same network, cooperative localization can relieve the requirement for multiple anchors. Furthermore, when devices are equipped with full-duplex transceivers, they can localize themself and map their surrounding environment with only a single RIS anchor. 
Finally, an extended list of open research challenges relevant to the proposed RIS-enabled seamless \ac{las} concept was presented, including the necessity for anchor deployment optimization and optimized resource allocation schemes, algorithmic and privacy issues, as well as the role of multi-functional RISs. 

\bibliographystyle{IEEEtran}
\bibliography{reference}

\begin{thebibliography}{10}
\providecommand{\url}[1]{#1}
\csname url@samestyle\endcsname
\providecommand{\newblock}{\relax}
\providecommand{\bibinfo}[2]{#2}
\providecommand{\BIBentrySTDinterwordspacing}{\spaceskip=0pt\relax}
\providecommand{\BIBentryALTinterwordstretchfactor}{4}
\providecommand{\BIBentryALTinterwordspacing}{\spaceskip=\fontdimen2\font plus
\BIBentryALTinterwordstretchfactor\fontdimen3\font minus
  \fontdimen4\font\relax}
\providecommand{\BIBforeignlanguage}[2]{{%
\expandafter\ifx\csname l@#1\endcsname\relax
\typeout{** WARNING: IEEEtran.bst: No hyphenation pattern has been}%
\typeout{** loaded for the language `#1'. Using the pattern for}%
\typeout{** the default language instead.}%
\else
\language=\csname l@#1\endcsname
\fi
#2}}
\providecommand{\BIBdecl}{\relax}
\BIBdecl

\bibitem{kisseleff2020reconfigurable}
S.~Kisseleff \emph{et~al.}, ``Reconfigurable intelligent surfaces for smart
  cities: {Research} challenges and opportunities,'' \emph{IEEE Open J. Commun.
  Soc.}, vol.~1, pp. 1781--1797, Nov. 2020.

\bibitem{tr38855}
\BIBentryALTinterwordspacing
``{3GPP TR 38.855 V16.0.0: Study on NR positioning support (Release 16)}
  (accessed on 10-{Feb}-2023),'' Mar. 2019. [Online]. Available:
  \url{https://portal.3gpp.org/desktopmodules/Specifications/SpecificationDetails.aspx?specificationId=3501}
\BIBentrySTDinterwordspacing

\bibitem{chen2022wi}
C.~Chen \emph{et~al.}, ``{Wi-Fi} sensing based on {IEEE} 802.11bf,'' \emph{IEEE
  Commun. Mag.}, vol.~61, no.~1, pp. 121--127, Jan. 2023.

\bibitem{wymeersch2020radio}
H.~Wymeersch \emph{et~al.}, ``Radio localization and mapping with
  reconfigurable intelligent surfaces: {Challenges}, opportunities, and
  research directions,'' \emph{IEEE Veh. Technol. Mag.}, vol.~15, no.~4, pp.
  52--61, Oct. 2020.

\bibitem{garcia2021tutorial}
M.~H.~C. Garcia \emph{et~al.}, ``A tutorial on {5G NR V2X} communications,''
  \emph{IEEE Commun. Surveys Tuts.}, vol.~23, no.~3, pp. 1972--2026, Feb. 2021.

\bibitem{tr38845}
\BIBentryALTinterwordspacing
``{3GPP TR 38.845 V17.0.0: Study on scenarios and requirements of in-coverage,
  partial coverage, and out-of-coverage NR positioning use cases (Release 17)}
  (accessed on 10-{Feb}-2023),'' Sep. 2021. [Online]. Available:
  \url{https://portal.3gpp.org/desktopmodules/Specifications/SpecificationDetails.aspx?specificationId=3806}
\BIBentrySTDinterwordspacing

\bibitem{FD_MIMO_arch}
G.~C. Alexandropoulos \emph{et~al.}, ``Full-duplex massive multiple-input,
  multiple-output architectures: {R}ecent advances, applications, and future
  directions,'' \emph{IEEE Veh. Technol. Mag.}, vol.~17, no.~4, pp. 83--91,
  Oct. 2022.

\bibitem{RISE6G_D51}
\BIBentryALTinterwordspacing
``{RISE-6G Deliverable 5.1: Control for RIS-based localisation and sensing}
  (accessed on 10-{Feb}-2023),'' Apr. 2022. [Online]. Available:
  \url{https://rise-6g.eu/Documents/LIVRABLES/RISE-6G_WP5_D5.1_Final.pdf}
\BIBentrySTDinterwordspacing

\bibitem{strinati2021reconfigurable}
E.~Strinati~Calvanese \emph{et~al.}, ``Reconfigurable, intelligent, and
  sustainable wireless environments for {6G} smart connectivity,'' \emph{IEEE
  Commun. Mag.}, vol.~59, no.~10, pp. 99--105, Oct. 2021.

\bibitem{tr38455}
\BIBentryALTinterwordspacing
``{3GPP TS 38.455 V17.2.0: NR Positioning Protocol A (NRPPa) (Release 17)}
  (accessed on 10-{Feb}-2023),'' Sep. 2022. [Online]. Available:
  \url{https://portal.3gpp.org/desktopmodules/Specifications/SpecificationDetails.aspx?specificationId=3256}
\BIBentrySTDinterwordspacing

\bibitem{keykhosravi2022ris}
K.~Keykhosravi \emph{et~al.}, ``{RIS}-enabled {SISO} localization under user
  mobility and spatial-wideband effects,'' \emph{IEEE J. Sel. Topics Signal
  Process}, vol.~16, no.~5, pp. 1125--1140, May. 2022.

\bibitem{Hyowon_RIS-SLAM_TWC2022}
H.~Kim \emph{et~al.}, ``{RIS-enabled} and access-point-free simultaneous radio
  localization and mapping,'' \emph{arXiv preprint arXiv:2212.07141}, 2022.

\bibitem{gu2022intelligent}
X.~Gu \emph{et~al.}, ``Intelligent surface aided {D2D-V2X} system for
  low-latency and high-reliability communications,'' \emph{IEEE Trans. Veh.
  Technol.}, vol.~71, no.~11, pp. 11\,624--11\,636, Jul. 2022.

\bibitem{ko2021v2x}
S.-W. Ko \emph{et~al.}, ``{V2X}-based vehicular positioning: {O}pportunities,
  challenges, and future directions,'' \emph{IEEE Wireless Commun.}, vol.~28,
  no.~2, pp. 144--151, Mar. 2021.

\bibitem{Tsinghua_RIS_Tutorial}
M.~Jian \emph{et~al.}, ``Reconfigurable intelligent surfaces for wireless
  communications: {O}verview of hardware designs, channel models, and
  estimation techniques,'' \emph{Intell. Converged Netw.}, vol.~3, no.~1, pp.
  1--32, Mar. 2022.

\end{thebibliography}


 




\begin{IEEEbiographynophoto}{Hui Chen}
(hui.chen@chalmers.se) is a postdoctoral researcher at Chalmers University of Technology, Sweden. \end{IEEEbiographynophoto}

\begin{IEEEbiographynophoto}{Hyowon Kim}
(hyowon@chalmers.se) is a postdoctoral researcher at Chalmers University of Technology, Sweden.
\end{IEEEbiographynophoto}

\begin{IEEEbiographynophoto}{Mustafa Ammous} (mustafa.ammous@mail.utoronto.ca)
is a Ph.D. student at University of Toronto, Canada.
\end{IEEEbiographynophoto}

\begin{IEEEbiographynophoto}{Gonzalo Seco-Granados} (gonzalo.seco@uab.cat)
is a professor at Universitat Autonoma of Barcelona, Spain.
\end{IEEEbiographynophoto}

\begin{IEEEbiographynophoto}{George C. Alexandropoulos} (alexandg@di.uoa.gr)
is an assistant professor at the Department of Informatics and Telecommunications, National and Kapodistrian University of Athens, Greece.
\end{IEEEbiographynophoto}

\begin{IEEEbiographynophoto}{Shahrokh Valaee} (valaee@ece.utoronto.ca)
is a professor at University of Toronto, Canada.
\end{IEEEbiographynophoto}

\begin{IEEEbiographynophoto}{Henk Wymeersch} (henkw@chalmers.se)
is a professor at Chalmers University
of Technology, Sweden.
\end{IEEEbiographynophoto}

\vfill

\end{document}